\documentclass[reprint,onecolumn,amsmath,amssymb,jap,aip]{revtex4-1}
\usepackage{graphicx}
\usepackage{dcolumn}
\usepackage{bm}
\usepackage{epsfig}
\usepackage{mathptmx}
\usepackage{times}
\usepackage{amsmath}
\usepackage{amssymb}
\usepackage{dcolumn}
\usepackage{units}
\usepackage{upgreek}
\usepackage{tabularx}
\usepackage{todonotes}
\usepackage{threeparttable}
\usepackage[space]{grffile}
\usepackage[normalem]{ulem}

\usepackage{subfig}
\usepackage{multirow}
\usepackage{array}
\usepackage{xcolor}
\usepackage{soul}
\usepackage[]{caption}
\begin{document}

\title{Phase Change Memtransistive Synapse}\affiliation{IBM Research -- Europe, S\"{a}umerstrasse 4, 8803 R\"{u}schlikon, Switzerland}
\author{Syed Ghazi Sarwat}\affiliation{IBM Research -- Europe, S\"{a}umerstrasse 4, 8803 R\"{u}schlikon, Switzerland}
\author{Benedikt Kersting}\affiliation{IBM Research -- Europe, S\"{a}umerstrasse 4, 8803 R\"{u}schlikon, Switzerland}
\author{Timoleon Moraitis}\affiliation{IBM Research -- Europe, S\"{a}umerstrasse 4, 8803 R\"{u}schlikon, Switzerland}\affiliation{Present Address: Huawei- Zurich Research Center, Zurich, Switzerland}
\author{Vara Prasad Jonnalagadda}\affiliation{IBM Research -- Europe, S\"{a}umerstrasse 4, 8803 R\"{u}schlikon, Switzerland}
\author{Abu Sebastian} \affiliation{IBM Research -- Europe, S\"{a}umerstrasse 4, 8803 R\"{u}schlikon, Switzerland}

\maketitle

\begin{flushleft}
\end{flushleft}

\noindent{\textbf{In the mammalian nervous system, various synaptic plasticity rules act, either individually or synergistically, and over wide-ranging timescales to dictate the processes that enable learning and memory formation. To mimic biological cognition for artificial intelligence, neuromorphic computing platforms thus call for synthetic synapses, that can faithfully express such complex plasticity and dynamics. Although some plasticity rules have been emulated with elaborate CMOS and memristive circuitry, hardware demonstrations that combine multiple plasticities, such as long-term (LTP) and short-term plasticity (STP) with tunable dynamics and within the same low-power nanoscale devices have been missing.  Here, we introduce phase change memtransistive synapse that leverages the non-volatility of memristors and the volatility of transistors for coupling LTP with homo and heterosynaptic STP effects. We show that such biomimetic synapses can enable some powerful cognitive frameworks, such as the short-term spike-timing-dependent plasticity (ST-STDP) and stochastic Hopfield neural networks. We demonstrate how, much like the mammalian brain, such emulations can establish temporal relationships in data streams for the task of sequential learning and combinatorial optimization.}}

\section{Introduction}

\noindent Sensory perception, motor learning or simply solving mathematical problems iteratively are some important cognitive tasks, where data streams are continuous and sequential. While the mammalian brain can process such sequential data effortlessly, artificial neural networks continue to only perform well with static, non-sequential information. Neuromorphic computing aims to bridge this gap, inspired by biological mechanisms. The powerful but efficient biological processing of sequential data involves a variety of synaptic mechanisms that are combined even within individual synapses, with dynamics that vary across a range of time scales. For example, using “non-volatile” plasticity (i.e. LTP)\cite{Prezioso2015,Bliss1993}, such as Hebbian-like or spike-timing dependent plasticity (STDP), neural circuit models and inspired neuromorphic algorithms can learn from event-based sequences~\cite{Nessler2013} and approximate deep learning~\cite{Y2017scellierequilibrium}. Note that here we use the acronym LTP for long-term plasticity including both long-term potentiation and long-term depression\cite{Citri2008}. On the other hand, “volatile” plasticity, i.e. STP, can act as a temporal filter\cite{Y2005grandesynaptic}, and a combination of STP with LTP enables learning on different timescales of sequential data. For the latter, yet another combination, namely STP with Hebbian plasticity in the form of ST-STDP, allows optimal adaptation to dynamic environments. Such mechanisms are simple, yet effective approach to difficult dynamic tasks, through which neuromorphic models have recently outperformed some deep neural networks\cite{Y2020Moraitisshort,Y2020Buonomano}. Therefore, it becomes evident that an ideal neuromorphic synapse for learning and inference should not only keep its size small, but also independently implement both LTP and STP, and be tunable to different input time scales. In addition, synapses should be able to adapt both individually and globally as in biological hetero~\cite{Regehr2012,Ohno2011a,Moraitis2018}- and homo\cite{Royer2003,Gkoupidenis2017}-synaptic plasticity. Recently, memristive nanodevices as synapses have emerged for neuromorphic computing. However, in these devices LTP and STP are tied together through a single governing mechanism. For example, in resistive memories both are dictated by the filamentation dynamics of atomic channels\cite{Y2016WangNatMat,Ohno2011a,Xia2019}, and the short-term dynamics are fixed and determined by the material. Thus, a new versatile and tunable synaptic nano device is needed for implementing high performing neural networks.\\

Here we demonstrate just such a device framework using “memtransistors”. Memtransistors are emerging analog devices, that exploit the properties of memristors and transistors. Recently they have been demonstrated using defect modulated Schottky barriers in two dimensional materials~\cite{Sangwan2018,Y2020Memtransistor}. In this paper, we illustrate the concept of phase change memtransistors, using the contemporary phase-change memory technology~\cite{Sarwat2017b,Sebastian2018}. Phase change memory is based on phase change materials, namely some selected chalcogenide glasses and metal dichalcogenides. The technology utilizes non-volatile, yet reversible and sub-nanosecond timescale phase transitions between at least two structural states of the material to store data (bits). The data is therefore encoded in the material’s tunable atomic configurations, each distinctive in its electrical conductance. Another property of most phase change materials is that they are semiconductors\cite{Y2006Liaocharacterization,Y2018Dausge2sb2te5}. The inherent semi-conductivity allows the electrical conductivity to also be tuned electrostatically, without requiring a change in the material’s atomic arrangement. This field-effect modulation is therefore contrastingly transient (volatile). We leverage these two independent features by creating a memtransistive synapse, whose non-volatility (from amorphous-crystalline phase transitions) enables LTP, and volatility (from electronic Fermi level shifts) enables transient plasticity. We show that such memtransistors can enable advanced neurosynaptic computations. Specifically, we demonstrate their use in the implementation of biomimetic neural algorithms for solving computationally difficult problems, namely sequential learning and stochastic Hopfield computing networks.

\section{Memtransistive Synapse}

\begin{figure}[h!]
    \includegraphics[width=\textwidth]{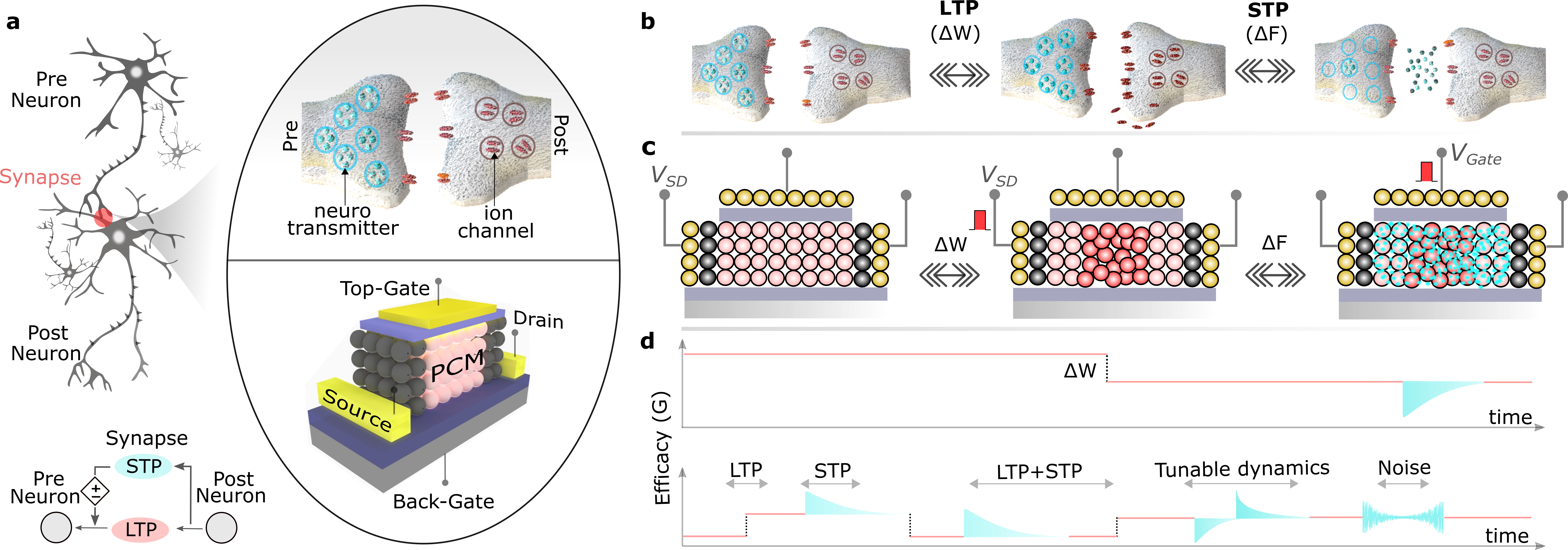}
    \caption{\textbf{Synaptic Efficacy and Phase Change Memtransistors}.(a) An illustration of biological neurons. The zoomed-in view is of a biological synaptic junction (synapse), and its artificial emulation. Synaptic efficacy ($G$) is dictated by the combination of the non-volatile long-term ($W$) and the transient ($F$) changes, from LTP and STP respectively. When a pre-synaptic spike from an input neuron creates a post-synaptic spike in the output neuron, $G$ may be transiently facilitated ($W+F$) or fatigued ($W-F$). A multi-terminal phase change memtransistor can adequately represent a synapse. The atomic-configurational changes between amorphous and crystalline states encode $W$, and the electrostatically modulated conductivity encode $F$. (b) An example representation of the biophysical mechanisms leading to the decrease in the non-volatile conductance ($\Delta W$) and volatile conductance ($\Delta F$), and corresponding bio-mimicry on the artificial synapse using the $V_{\text{SD}}$ and $V_{\text{Gate}}$ signals. (d) Time-dependent $G$ evolution in a synthetic synapse. The top panel depicts how $G$ gets modulated when $W$ and $F$ are changed for the case illustrated in (c). The bottom panel highlights the various synaptic processes a memtransistive device can emulate, including both non-volatile and volatile increases and decreases in $G$, noise, and tunable relaxation dynamics.} 
    \label{Fig:1}
\end{figure}

\noindent When both the LTP and STP effects are considered, the synaptic efficacy ($G$), or how strongly neurons connect, can be expressed as $G(t) = W(t) \pm F(t)$. In this expression, $W(t)$ denotes the stored synaptic weight governed by LTP, and $F(t)$ is a decaying term governed by transient plasticity (see Figure \ref{Fig:1}a-b). With this expression, $G(t)$ thus captures a range of neurosynaptic processes: LTP, STP, LTP $\pm$STP, transient noise, and tunable dynamics, each of which has important implications in cognitive workloads. For example, with tunable dynamics, the change in synaptic strength can scale disproportionately in time (ms to minutes to days). This allows implementing tasks such as reward-based learning that require persistent retention~\cite{Gerstner2018}, as well as sensory perception, which requires quick forgetting. Our phase-change memory memtransistive devices can capably emulate all of the above synaptic processes (see Figure \ref{Fig:1}c-d).\\

\begin{figure}[h!]
    \includegraphics[width=0.9\textwidth]{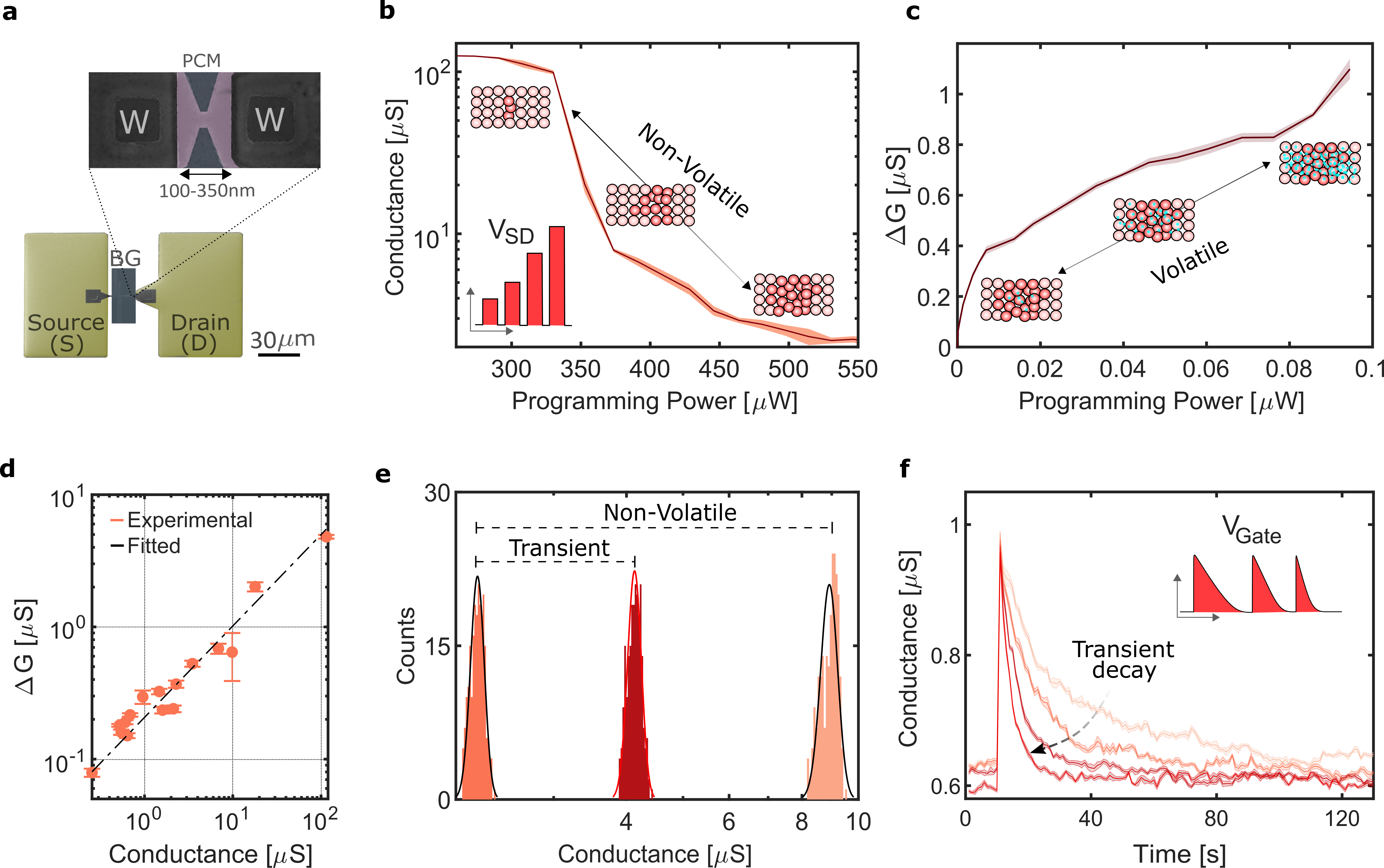}
    \caption{\textbf{Dynamics of Memtransistive Synapses}.(a) False colored scanning electron micrograph of a lateral line cell, with phase-change material as the active channel. (b) Programming curves of a memtransistive synapse. The device is reversibly switched multiple times between many non-volatile conductance states using source-drain signals ($V_{\text{SD}}$), each encoding a unique $W$. (c) Field-effect programming curves of a memtransistive synapse in an arbitrary $W$ state. The device is modulated multiple times between many volatile conductance states using gate-drain signals ($V_{\text{Gate}}$), each encoding a unique $F$. (d) A plot illustrating the dependency of $F$ on $W$. $F(W)$ shows a positive relation, that can be adequately fitted (black trace) with an exponential function. (e) A histogram plot highlighting the stochasticity in the conductances of the programmed non-volatile, and volatile states in a nano-cell. (f) A plot illustrating the tunable relaxation dynamics of a transient state.}
    \label{Fig:2}
\end{figure}

\noindent The demonstrator memtransistors studied here are cells of $Ge_{\text{1-x}}Sb_{\text{x}},\,x=[1,0.85]$ on $Al_{\text{2}}O_{\text{3}}$ and $SiO_{\text{2}}$ dielectrics, wired in a 4-terminal configuration. We investigated two different device configurations: a nano and a micro-cell (see Figure \ref{Fig:2}a and Section S1, respectively). Both are lateral types but differ in dimensions (channel length of sub-350 nm and sub-10$\mu m$ for the nano and micro cell, respectively). The devices can be programmed to different non-volatile conductance states using Joule heating from write and erase pulses, sent across the source-drain ($V_{\text{SD}}$) terminals, as is illustrated in Figure \ref{Fig:2}b for a nano-cell (the measurement is repeated 6 times in this plot). These analog non-volatile conductance states are representative of LTP. The transient changes in the electrical conductance are enabled by voltage pulses, applied across the gate-drain terminals, as is shown in Figure \ref{Fig:2}c (the measurement is repeated 10 times in this plot). The devices can be programmed to many transient conductance states using the electrostatic modulation of the channel. These volatile conductance states are representative of STP. The global transient effects, such as homeostatic plasticity are triggered using the bottom-gate ($V_{\text{BG}}$), which is common to all the devices on-chip. And the local transient plasticity such as STP is implemented using the top-gate ($V_{\text{TG}}$), unique to each device, or through the back-gate (however, with routed and asynchronous/jittered inputs; see section S3). The field-effect and temperature-dependent measurements show that the devices switch from ambipolar to \textit{p}-doped degenerate semiconductors, when they fully crystallize (see section S2). Additionally, the extent of transient change in the channel's conductance ($\Delta G$ or $F$) from a gate-drain voltage signal is observed to scale with the device conductance ($W$), as shown in Figure \ref{Fig:2}d for a $100$ nm length $Sb$ nano cell (the measurement is repeated 10 times in this plot). The higher conductance states are electrostatically more responsive (transconductive), and akin to biological synaptic strength scaling, $F(W)$ shows an exponential relationship (shown by the adequate fit in Figure \ref{Fig:2}d). To provide clarity on the rather novel neuromorphic computations that we will discuss, we limit our memtransistive devices to operate under a binary mode (to toggle between two conductance states). In Figure \ref{Fig:2}e a typical programming experiment is plotted, where a $175$ nm long nano cell is reversibly switched between two conductance states. Each state is non-volatile and addressable using a pulse programming scheme. Also plotted is the programming noise of the transient state achieved when the lower conductance state is electrostatically modulated with a gate signal (-20 $V_{\text{BG}}$). The normal distribution of these conductance histograms indicate that both the non-volatile and volatile states are stochastic, much like in biological synapses~\cite{Neftci2016,Tuma2016}. For tunable synaptic dynamics, the synapses should be programmable for the relaxation (decay) rate of its transient state, in ways that do not affect the baseline non-volatile states. Figure \ref{Fig:2}f illustrates such a tunable behavior, where a transient state is programmed by $V_{\text{BG}}$ pulses of the same amplitude but varying decay rates (the measurement is repeated 10 times in this plot). For such pulses the device conductance evolve as $F = F_0\exp(-{\alpha t})$ ($F$ is instantaneous conductance, $F_0$ is maximum conductance, $\alpha$ is the relaxation rate and $t$ is time). Notably, the volatile state of a memtransistive cell can be adjusted to relax at varying rates to suit the activity of the task at hand or for a different task altogether.

\subsection{Local plasticity for sequential learning}

\begin{figure}[h!]
    \includegraphics[width=\textwidth]{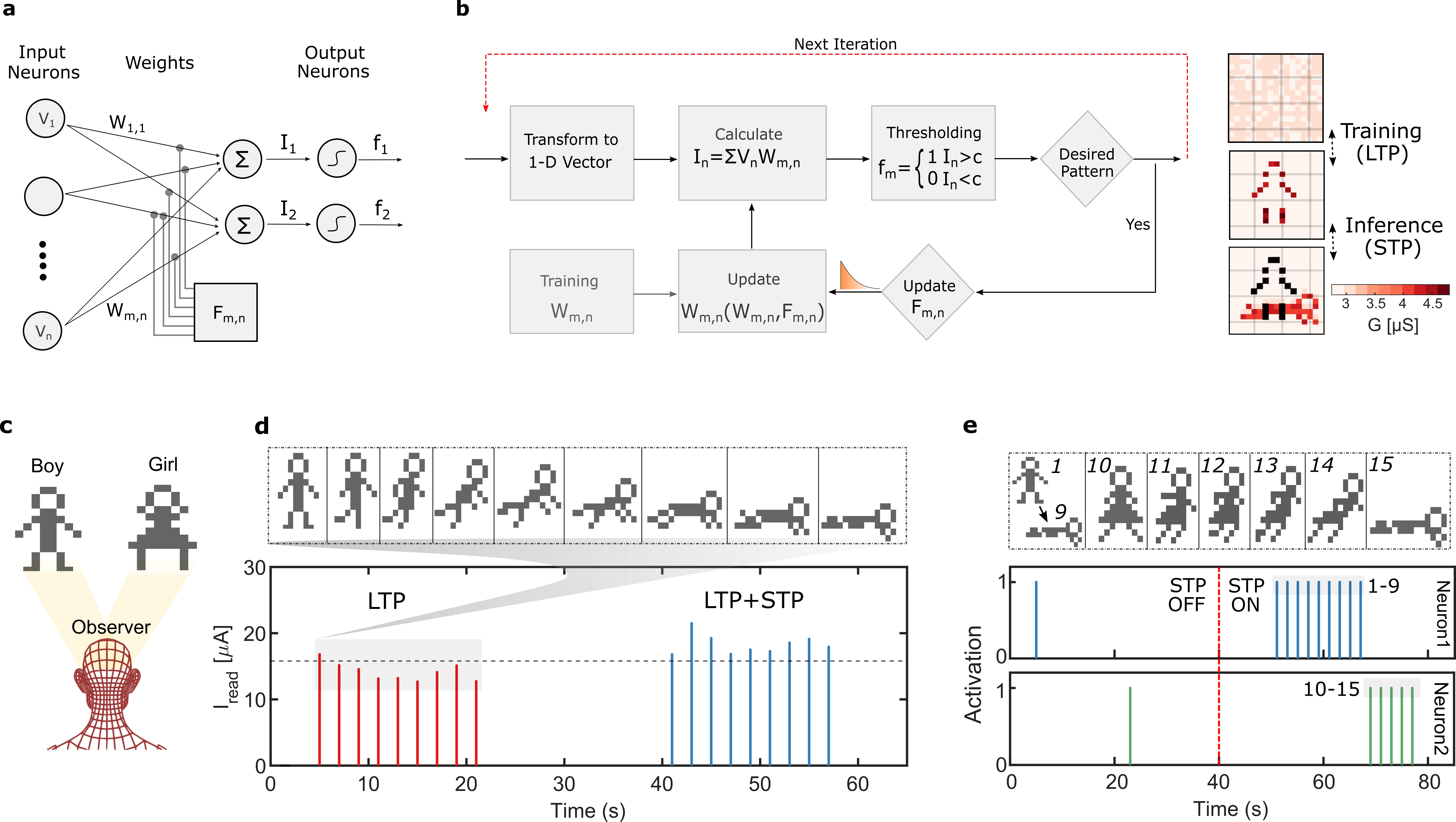}
    \caption{\textbf{Recognizing sequential data with short-term spike dependent plasticity}.(a and b) An illustration of a single-layer neural network and a flow chart describing the key computational operations. The weights ($W_{\text{m,n}}$) are learned through the non-volatile LTP effects. The right most panel in (b) shows conductance matrices of the network before and after training with LTP, and during inference with STP.  (c) Cartoon illustration of the performed classification task. The receptive field of the observer (a computer) allows it to intermittently visualize either the boy or girl, both of which morph with time. (d) A plot illustrating the summed output current ($I_{\text{read}}$) of a trained memtransistive array to an input of moving images (sequentially changing image frames of the boy), with only LTP (red trace), and with LTP and STP (blue trace) effects. (e) Classification of the image frames associated with both sequential and non-sequential data streams (top panel). The memtransistive array can capably recognize the sequential data (body movements) as well as distinguish distinctive data (boy and girl).}
    \label{Fig:3}
\end{figure}

\noindent The ability to use the information of a past event to predict a current or future event lays the premise for sequential learning. We carry out the task of classifying running image frames by using a single-layer neural network (classifier). As illustrated in Figure \ref{Fig:3}a, the classifier comprises an input layer, consisting of input neurons to which the sensory data  (as voltage pulses ($V_n$)) is fed, a synaptic layer, where the trainable weights ($W_{\text{m,n}}$) are encoded in the non-volatile states of the memtransistors, a summation layer, an activation ($f_m$) layer and finally an output neurons layer. Uniquely to this classifier, we add a transient plasticity unit ($F_{\text{m,n}}$) that temporally tunes the conductance of select synapses. A flowchart that describes the network is presented in Figure \ref{Fig:3}b.  The network is trained using a backpropagation scheme. The memtransistors carry out the vector-matrix multiplication operation on a training data set ($I_n=\sum{W_{\text{m,n}}V_{\text{n}}}$, where $I_i$ is the output current). After training, the weights are updated and the network learns to classify static inputs.  During inference, input that is either similar to or contrastingly different from the training data is presented to the network. Crucially, it is during inference when the transient plasticity is utilized, and temporal relationships between the data streams are learned. Figure \ref{Fig:3}b right panel illustrates the conductance states of all the memtransistive synapses in the network before and after training, and during inference. LTP dominates the training process and selectively heightens the conductance of select synapses, which do not change later (until otherwise the  network is retrained). While during inference, STP heightens the conductances of only those synapses that correspond to the input. This facilitation is short-lived, and the heightened conductances gradually relax to the trained baseline states. \\

The task we perform is to classify a boy and a girl. The observer (computer vision) can visualize either the boy or the girl at any time instance and both these objects (features) can morph in time (Figure \ref{Fig:3}c). Figure \ref{Fig:3}d illustrates the case when the observer’s vision is receptive to only the black and white image frames of the boy, who is falling sidewards. Nine distinct and yet associable image frames of the boy are input to the network, in the sequence shown in the top panel of Figure \ref{Fig:3}d. The input is presented in two batches (each batch comprises nine sequential frames). In the first batch, the network is required to predict the frames only with LTP, while in the second, the transient STP is triggered to complement LTP. Note that with LTP (red trace) the network’s output is greatest for the first frame (beyond the network’s classification threshold),  while for the successive frames it diminishes. This highlights the inability of the network to associate the subsequent frames to the boy and is a result of the decreasing overlap between the input features and the trained weights. For LTP to perform this task, the network must be trained with all possible transformations the boy could make. However, this approach is impractical and a bottleneck, for there can exist an infinite number of transformations. STP (blue trace) aids the task by establishing relationships between the moving frames, using the first frame that is closest to the trained weights as a trigger (reference). This is a result of synaptic facilitation or the transient increase in the synaptic conductances following the successful recognition (post-synaptic spike) of the first frame. Figure \ref{Fig:3}e further elaborates this. We plot the state of the two output neurons to a data set comprising input frames of the falling boy and girl. The experiment is iterated twice in the figure: with and without STP. The sequentiality in the data is represented in the body movements of both boy and girl, while the discontinuity is encoded in the distinctive nature of the frames (boy vs girl). Thus, only the first nine (1-9) and the following five frames (10-14) are temporally related in the task. Without STP, the neurons corresponding to boy and girl fire only for the first frames. With STP, note that output neuron 1 which corresponds to the boy is activated, while neuron 2 which corresponds to the girl is dormant when frames 1-9 are presented to the network. The neuronal activations reverse when the girl’s frames are presented and the network also capably establishes temporal relationships in the girl’s sequence. Also, note that frame 15 (also frame 9 in the boy’s sequence) when appearing after the girl’s frame does not get recognized by either neuron, since it is not sequential. This ability of the network to see through the sequential frames, and yet distinguish them highlights an important attribute of STP- that is the utility of past information does not comprise classification of new data sets.\\

\begin{figure}[h!]
    \includegraphics[width=\textwidth]{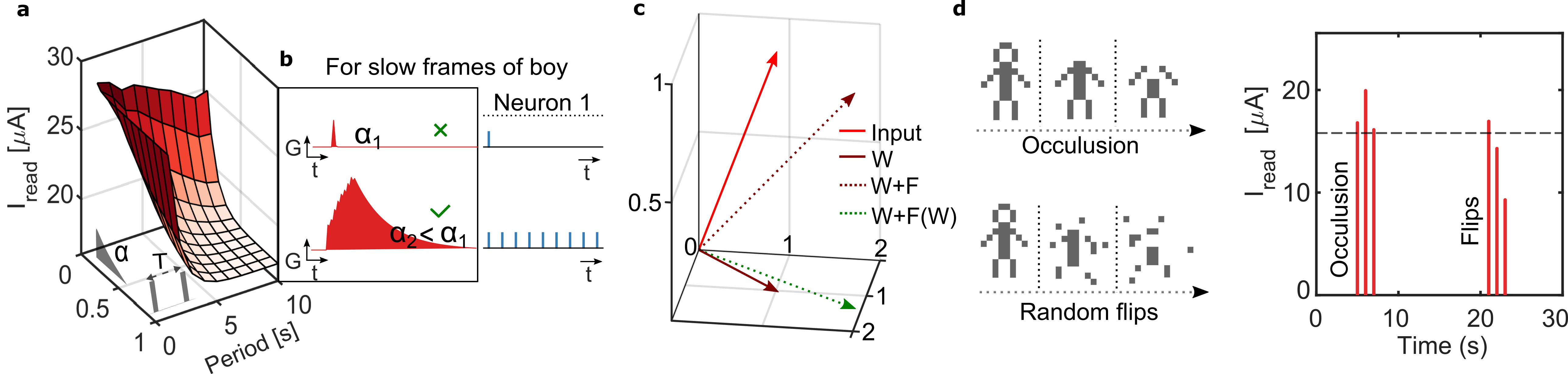}
    \caption{\textbf{Tunable and weight-dependent short-term dynamics}. (a) A graph illustrating the inverse dependency of the array’s output on the rate (x-axis) at which frames are perceived, and the decay time of the synaptic transient effects (y-axis). (b) The summed conductance ($G$) of the memtransistive array when images of the boy from Figure \ref{Fig:3}d are perceived under two different relaxation constants ($\alpha$). (c) A conceptual illustration of sequential learning with an illustrative example of three synaptic weights. (d) An example illustrating the use case of $F(W)$ for two data sets, one where the boy's frames are sequentially occluded, and the other where the pixels are randomly flipped.} 
    \label{Fig:4}
\end{figure}

The long-term and transient plasticity of the synapses provide compensatory effects. This to an extent that the transient plasticity -which requires no prior training- can enable accurate inference even for inputs that only partially overlap with the trained weights (see section S4). In Figure \ref{Fig:4}a we illustrate the use of tunable dynamics for recognizing inputs with varying frames rates. The panel plots the dependence of $I_{\text{read}}$ on both the rate at which information is perceived, and the synaptic relaxation constants. Markedly, the frame rate (waiting period between the inputs) can be compensated with the decay constant, such that depending on the application, both fast and slow-moving input frames can be sequentially recognized. Figure \ref{Fig:4}b illustrates an exemplar case (also see section S4), in which we plot the $G$ of the network only when an input frame in the boy's nine slow moving frames (see Figure \ref{Fig:3}d) is detected. When a faster decay constant ($\alpha_{1}$) is used, only the first frame is recognized, but when the decay constant is adjusted to ($\alpha_{2}<\alpha_{1}$), past inputs aid in the recognition of present inputs and all image frames are recognized (depicted by the spikes in output Neuron 1).\\

In natural environments, most transformations are sequential, except when an irregularity is imposed. The computer vision abruptly turning its attention  from boy to girl, is an example of the latter. In some cases, it therefore becomes important to constrain the synaptic efficacies change. This is achievable if the transient plasticity itself is state-dependent, such that $G=W+F(W)$. Figure \ref{Fig:4}c illustrates this with using euclidean space, with three synapses ($W_{1},W_{2},W_{3}$), such that the vector $[1\,1\,0]$ (maroon solid line) would represent the trained network. When an input vector $[1\,1\,1\,]$ (red solid line) is presented to the network, ST-STDP ($G(t)=W(t)+F(t)$) introduces both facilitation (increase in vector length), and a change in direction, such that the synaptic weights are attracted to the input (maroon dotted line).  If $F$ scales with $W$, in the specific case that $F(W)$ is positively correlated, then $G(t)=W(t)+F(t,W)$ constrains the direction change (green dotted line), but not the facilitation. In the purview of visual information, this implicates that an object can only be a transformed version of itself. In Figure \ref{Fig:4}d, we illustrate this with an example of image occlusion and random pixel flips on the object boy. Only occlusion  represents a continuously transforming environment. What is desired here is for the network to recognize the occluded frames and not the random flips. The right panel of the figure plots the $I_{\text{read}}$ for this data set. Notably, with ST-SDTP, sequential frames under occlusion are recognized, but not under random flips. It is also noteworthy that, because $F(W)$ is a positive function, sequential learning can be also realized with a global gate-modulation since only higher conductance states will get facilitated most. In passing, another prospect of using the ST-STDP framework could be to utilize it as an indeterminate component in other networks. For instance, the inverse relationship of $I_{\text{read}}$ with the input delay and the relaxation dynamics can be utilized for the task of classifying events, such as an object's motion (or activity). In section S4 we discuss the classification of motion in the boy's frames.

\subsection{Global plasticity for synaptic scaling}

\begin{figure}[h!]
    \includegraphics[width=0.9\textwidth]{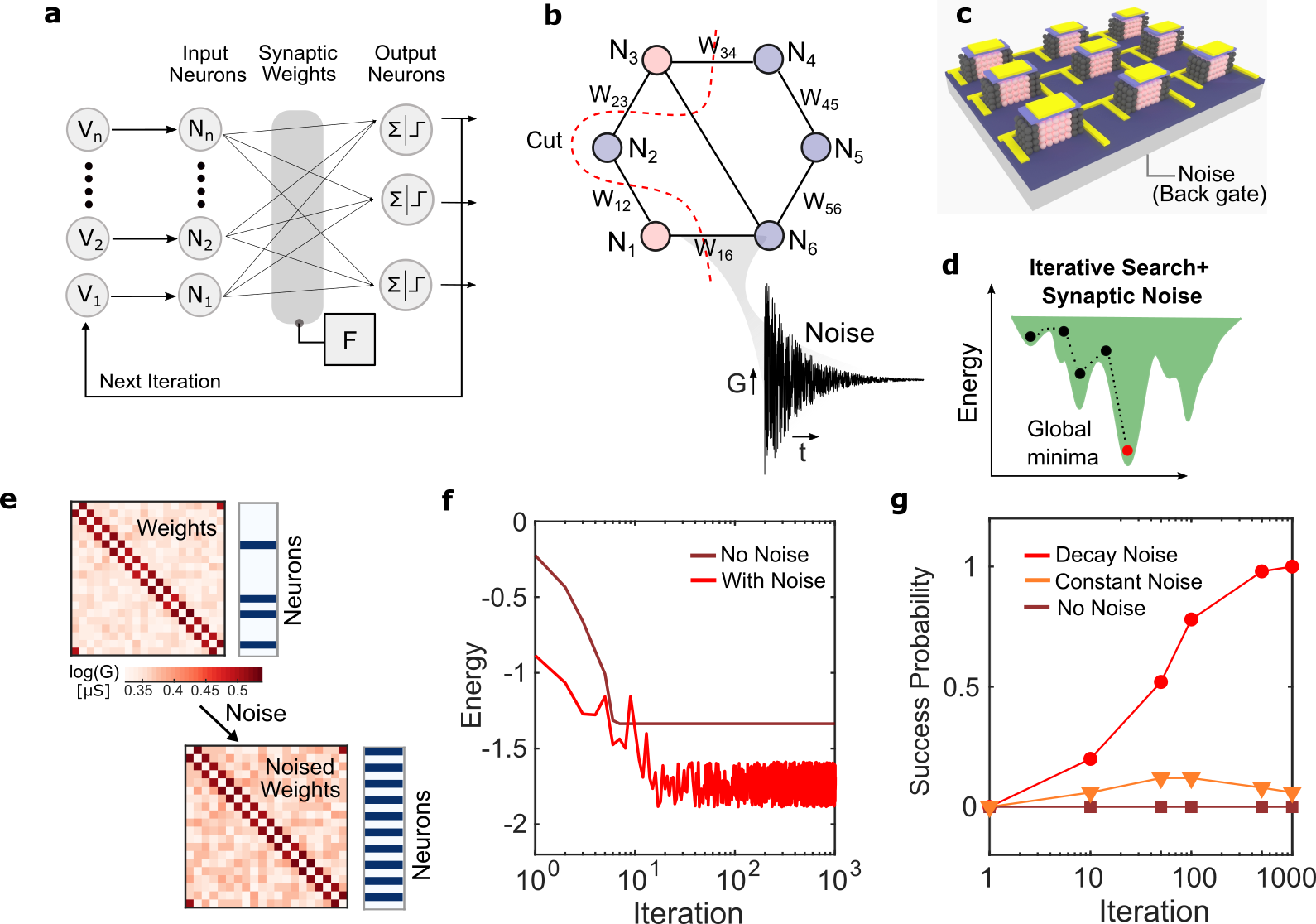}
    \caption{\textbf{Combinatorial optimization using global plasticity}. (a) A schematic illustration of a recurrent Hopfield network. (b) A cartoon of a graph optimization problem of Max-cut. The dashed red line cuts the graph into two sets of two complementary nodes. The edges are encoded as non-volatile synaptic weights and the nodes as neurons. (c) A cartoon illustration of memtransistive synaptic weights and the use of a global back-gate for noise injection. (d) The energy landscape of a typical NP-hard problem, where the black dots are representative of the system state.  (e) An illustration of the conductance distribution of the non-volatile memtransistive-synaptic weights, and an initial distribution of the neuronal states. The bottom panel highlights the temporal modifications to the weights from added synaptic noise, and through it a convergence towards the optimal distribution of the neuronal states. (f) The Hopfield energy of the system with and without decaying synaptic noise. (g) A comparison of the effect of decaying noise, a fixed noise, and no noise on the statistical success of the system to convergence.}
    \label{Fig:5}
\end{figure}

\noindent So far we discussed the transient plasticity effects, subject to select synapses, that were resultant of the homosynaptic effects. Some transient mechanisms are homeostatic, such as hormonal and temperature influenced homeoplasticity effects. These modulate the synaptic efficacies globally ~\cite{Fernandes2016,Watt2010,Bains2014}, i.e. at any time instance a network of synapses are transiently affected.  We now show that memtransistive synapses can emulate such homeoplasticity effects, and through it enable hardware accelerators for computationally difficult problems. Examples of such problems are combinatorial optimization 
, which are NP (non-polynomial) hard\cite{Y2010Combitroialcomplexity,Y2011combinatorial}. Figure \ref{Fig:5}a shows a bioinspired Hopfield neural network (HNN), comprising phase change memtransistors as synaptic weights. The network is re-current (i.e. every processing neuron unit is connected to all other neurons), and the transient plasticity unit ($F$), when triggered using a global back gate influences all synapses simultaneously. For demonstration, in such a network we encode the constrained optimization graph problem of Max-Cut (see Figure \ref{Fig:5}b). Max-cut is typically utilized in integrated-circuit designing and imaging, and the goal is to partition a graph's vertices into two complementary sets (highlighted in different colors), such that the number of edges linking the two sets is as large as possible\cite{Y2003MAXCUT,Y2011combinatorial}. On the hardware, the edges are represented by the memtransistive synaptic weights $W_{\text{m,n}}$ (see Figure \ref{Fig:5}c). The problem is mapped to an energy function\cite{Y2001DISCRETEHOPFILED} $E$ (see Methods). \\

Such an HNN framework converts all problem constraints into solution objectives and thereby produces multiple end solutions. This is illustrated in the energy landscape cartoon shown in Figure \ref{Fig:5}d.  The solutions corresponding to the global minima are of interest (optimal convergence) .  Among other reasons, however, when such problems are solved, the solutions most often are stuck at the parasitic local minima. This is typically overcome using probabilistic (uncorrelated) techniques, such as simulated annealing. We find that the memtransistive synapses with homeoplasticity rendered conductance fluctuations are perfectly suited for enabling simulated annealing. In this approach, $W_{\text{m,n}}$ is a function of $F(t)$:  $W_{\text{m,n}}(t)=W_{\text{m,n}}\pm F(t)$, where $F(t)$ is synaptic noise. In Figure \ref{Fig:5}e, we find Max-cut in an undirected cyclic graph with 20 vertices. The memtransistive-synaptic weight matrix corresponding to such a graph is illustrated. The system is initialized with a random distribution of the neuronal states, depicted as a vector in the figure, where the dark shade is representative of the neuron belonging to set \text{S$_{1}$}, while the light shade is indicative of the neuron belonging to the complementary set \text{S$_{2}$}. When the optimization is performed, the system converges to the neuronal state distributions that minimize the net energy. When synaptic noise is added to the system, the synaptic-weight distributions are altered (transiently), and the system converges faster to the optimal solution. This is illustrated in the bottom panel of Figure \ref{Fig:5}e; note that the neuronal distributions have changed such that they represent a Max-Cut (most number of elements/edges in either set). In Figure \ref{Fig:5}f, we plot the Hopfield energy of the network with and without synaptic noise. Without noise injection (brown trace), the network converges to a sub-optimal solution within a few iteration cycles, and then no longer changes. This state corresponds to a local minimum. When the stochastic synaptic noise is added, the energy of the network fluctuates and thereby gets lowered (red trace). The neuronal configuration corresponding to this energy corresponds to the optimal solution. \\

We note that the magnitude and duration of the conductance fluctuations are an important consideration in realizing a statistically significant energy reduction. The memtransistive synapses uniquely have a control mechanism that we previously discussed for both noise magnitude and duration adjustments. This occurs through the modulations of $F_{\text{0}}$ and $\alpha$. To illustrate this, we compare the statistical success to convergence under no synaptic noise ($F_{\text{0}}=0$), constant synaptic noise ($F_{\text{0}}\neq 0$, and $\alpha=0$), and decaying noise ($F_{\text{0}}\neq 0$, and $\alpha \neq 0$). In Figure \ref{Fig:5}g, we plot the probability of success for each case, whereby the experiment is repeated 50 times for every iteration. Notably, the network fails to converge without noise, and under decaying noise, the success probability increases with an increasing number of iterations. However that under constant noise (or high noise), the convergence success is lowered. This is a result of the high magnitude and frequency of the fluctuations, which displace the system significantly from the energy minima. Thus, this model example illustrates the significance of the tunable dynamics of memtransistive synapses and their ability to emulate homeoplasticity effects for neuronal computations. We also note that such homeoplasticity effects can be utilized to maintain desired firing rates and patterns in the network. For example, in response to low or high neuronal activity, the network may enhance or decrease the excitability of the neurons to restore stability and accuracy\cite{Watt2010,Bains2014}, such as in the context of adaptive learning (see section S5). 

\section{Discussion}

\noindent Our approach of using memtransistive synapses for representing biological neural networks enables the emulation of some advanced dynamics of neural processes. The ability to co-locate the non-volatile and transient weight updates at the level of synapses provides savings in important computational metrics against larger recurrent and long-short term memory neural networks~\cite{Park2019, Dietterich2002}, where learning occurs on neuronal scale. Similarly, such a physics-based analog system allows for a scalable and efficient hardware implementation of a Hopfield network, which shows potential for a more efficient of execute complex optimization algorithms. We note that more recently,  memristive crossbars have been exploited for HNNs\cite{Y2020CaiNatureElectronics,Y2020TransientChaosScienceAdvances,Y2017SuhaseNbO2chaotic}. However, they do not exploit the otherwise rich plasticity of the biological synapses and thus require additional devices or sophisticated circuitries for combinatorial problems. \\

We also note that from a circuit-theoretical representation, a memtransistor can be represented using a combination of a field-effect transistor and a memristor, or could also be emulated using CMOS, similar to how memristors have been \cite{Y013CMOSEmulator}. The former approach could be more attractive since memristors already incorporate access transistors in various topologies, such as active crossbars. Some distinct features of memtransistors however do make them unique, such as the transient plasticity dependency of LTP. To be able to capture such dependencies on emulated platforms would require the use of additional storage buffers, making otherwise complex hardware, even more space expensive.  Similarly, the homeoplasticity effects can be more simplistically be realized, and importantly can be delineated from the rigid CMOS circuitry, operating at some hard-coded parameters. Nonetheless, some challenges also arise when discussing large-scale integration of the memtransistive synapses, since the need for multiterminal control calls for a rethinking of the peripheral circuitries and the crossbar layout (see section S3). Also a strong electrostatic effect in memtransistive materials becomes a requirement, due to the strict output voltage bound of the CMOS peripheral circuitries. Lastly, we also demonstrate that memtransistive synapses can be implemented on a charge-trap memory technology (see section S6), which expand our concepts to other material device systems, including silicon. Overall, our demonstrations hint at a new breed of bio-inspired device technology for implementing neuromorphic computing, as well as for testing ideas in neuroscience.

\section*{Data availability}

\noindent The data that support the findings of this study are available from the corresponding author upon reasonable request.

\section*{References}
\def\url#1{}

\section*{Methods}
\noindent\textbf{Device fabrication}: The devices were fabricated on Si wafers, that were capped by a 40 nm thermally grown $SiO_2$ dielectric layer. The bridge cell devices comprised a stack of 3 nm Sb and 5 nm $SiO_2$ thin films that were sputter deposited as blanket films at pressures 3 $\mu bar$ and 6 $\mu bar$, respectively. E-beam lithography was used to define the geometry of the bridge cells, using hydrogen silsesquioxane (HSQ) resist. After patterning, the blankets films were etched using Ar ion-milling. To avoid side wall oxidation, the cells were passivated with an 18 nm sputter deposited $SiO_2$, immediately after ion-milling. In order to make electrical contact with the cells, the capping layers at the extreme ends were locally opened using Ar ion-milling process in a second e-beam lithography. After stripping the resist, 60 nm W is sputter deposited. In yet another e-beam lithography step, W was patterned and etched at RIE to define contact electrodes and series resistor to the bridge cells. The chip was encapsulated with an 80 nm sputter deposited $SiO_2$ layer. Finally, Au probe pads of  200 nm thickness were sputter deposited, to make contact with the extended W structures.\newline 

\noindent{\textbf{Electrical characterization}: The electrical measurements on the nano cells were performed in a cryogenic probing station (JANIS ST-500-2-UHT) at variable ambient temperatures (100 K to 250 K). The temperature of the chuck was measured using a Lake Shore Si DT-670B-CU-HT diode. A Lake Shore Model 336 automatic temperature controller, and two heaters at the sample holder and chamber radiation shield controlled the temperature with an accuracy of 0.5 K. The measurements were performed under high vacuum, to the order of $10^{-6}$ torr to minimize heat exchange via convection and avoid water condensation. Devices were electrically contacted with a high-frequency Cascade Microtech Dual-Z GSSG probes. The probe was thermally connected to the sample holder with cooling braids. DC measurements of the device state were performed with a Keithley 2600 System SourceMeter. AC signals were applied to the device using an Agilent 81150 A pulse function arbitrary generator. A Tektronix oscilloscope (DPO5104) recorded the voltage pulses applied to and transmitted by the device. Switching between the circuit for DC and AC measurements was done with mechanical relays. Microcells were measured under ambient room conditions on a Cascade Probestation using a Keysight semiconductor parameter analyzer.}\newline

\noindent{\textbf{Sequential Learning}: In the implementation of STDP for the moving images, the synaptic weights ($W_{\text{m,n}}$) are updated with a decaying signal ($F_{\text{m,n}}$). Only following the image frames that the network correctly recognized, i.e. the output current ($I_\text{read}$) from vector-matrix multiplication between the input and trained weights exceeds some threshold. When triggered, $F_{\text{m,n}}$ modifies $W_{\text{m,n}}$ in the short-term, and thereby influences the vector-matrix multiplication operations of the network on the subsequently arriving image frames. Importantly, the degree to which $I_{\text{read}}$ of a subsequent image frame (pre-synaptic) is influenced by $F_{\text{m,n}}$ scales inversely with how far in time this frame follows the preceding frame (post-synaptic). These attributes are reminiscent of the spike-timing-dependent plasticity, conventionally utilized for non-volatile updates in spiking neural networks, however, here they appear as transient updates that are short-lived, therefore STP. Each pixel in the input frame is mapped onto either a high (0.2 V) or a low (0 V) voltage signal, depending on its intensity, and the mapped vector is presented to the network. The experiment for image recognition is performed for a defined time and the network’s $I_{\text{read}}$ is computed at each time instance. The network outputs a spike $I_{\text{read}}$ only for input frames that have some overlap with the trained weights, such that the product of at least one or many vector-matrix multiplication operations is non-zero. Therefore, the amplitude of $I_{\text{read}}$ scales proportionally with the number of non-zero vector-matrix multiplication operations. To perform above, we extracted experimental data from multiple nanocells at 200 K and used the data to simulate the learning task on Matlab. In applications relating to motion or activity sensing, the task is often to not disregard information with some threshold, neither it is to classify every time instance, instead, it is to monitor information over some defined period. To demonstrate this, we feed dynamically changing image frames of a boy and a girl to the network. The network starts by classifying a frame into an expected class (boy or girl), after which the transient plasticity in the synapses sets in, affecting the $I_{\text{read}}$ from the subsequent frames that follow, thus detecting the object’s activity. To classify the activity in a defined period, we perform a running average over every $I_{\text{read}}$ the network outputs. If the frames change more rapidly, the averaged output($I_{out}$) is larger than when they change less quickly, due to the decaying nature of the transient plasticity effects, as was discussed before. By mapping the magnitude $I_{out}$ into three categories, the network can classify the object's motion.}\newline

\noindent{\textbf{Combinatorial Optimization}: In solving the problem of Max-cut using a Hopfield neural network, we map the objective function of the graph problem to the Hopfield energy $E=-\frac{1}{2}\sum_{\text{mn}}^{\text{n}} W_{\text{m,n}}\nu_{\text{m}}\nu_{\text{n}}$, where $\nu_{\text{m}}$ and $\nu_{\text{n}}$ represent the state of interconnected neurons in a network of $N$ neurons, and take values of the read voltage (0.2 and -0.2 V). The energy has an arbitrary unit scale, which is a function of the graph problem (connection density and weight values), and the values of the neuronal states. To find an optimal solution, the energy function is iteratively minimized by toggling the neuronal states asynchronously, using the update-rules: $I_{Read\,(m)}=\sum_{m\neq n}^{}W_{\text{m,n}}\nu_{\text{m}}$, where $\nu_{\text{m}}$=+0.2 V if  $I_{Read\,(m)}\geq \theta_{\text{ref}}$ or $\nu_{\text{m}}$=-0.2 V if  $I_{Read\,(m)}<\theta_{\text{ref}}$  ($I_{Read\,(m)}$ is the read-current, where $\theta_{\text{ref}}$ is a threshold).  To perform above, we extracted experimental data from multiple nanocells at 200 K and used the data to simulate the learning task on Matlab. In each iteration, all neurons are updated (but asynchronously). However in another scheme each iteration could be associated with a randomly picked neuron and its updating. In our chosen graph problem, each vertex (neuron) has edges with only immediate neighbors ($W_{\text{mn}}=W_{\text{nm}}$), and there are no self-loops ($W_{\text{mn}}=0$, if $m=n$).  These devices (synapses) are programmed using $V_{\text{SD}}$ pulses, and the high-conductance memtransistive devices encode an edge (red shade in Figure \ref{Fig:5}e), and low-conductance devices represent the absence of edges (skin shaded in Figure \ref{Fig:5}e). The neuronal configuration in Figure \ref{Fig:5}e is such that the dark (blue) shade is representative of the positive read voltage (0.2 V) state, while the light (blue) shade is indicative of the negative read voltage (-0.2 V) state. Simulated annealing requires that the fluctuations are stochastic (uncorrelated) and that they decay as the system approaches the global minimum. We achieve this through  $F(t) = F_0exp^{\alpha t}$ (where the terms have their usual meanings). The transient plasticity can dislodge the system from local minima if stuck since $\nu_{\text{m}}$ can fluctuate during optimization. The control over $F(t)$ means that the system can be displaced from the low energy barriers defining local minima, but not from the high barriers defined global minima.}\newline 

\section*{Acknowledgments}
\noindent We acknowledge funding for this work from the European Union’s Horizon 2020 Research and Innovation Program (Fun-COMP project $780848$). A.S. acknowledges support from the European Research Council through the European Union’s Horizon $2020$ Research and Innovation Program under grant number $682675$. S.G.S acknowledges the stimulating discussions with Abbas Rahimi (Research Scientist, IBM Research-Zurich) on HNNs.

\section*{Competing financial interests}
\noindent The authors declare no competing financial interests. This work not done in collaboration with Huawei.\newpage

\end{document}